\documentclass[traditabstract]{aa}  
\usepackage{graphicx}
\usepackage{txfonts}

\begin{document}

\title{Hot subdwarf stars in close-up view}

\subtitle{IV. Helium abundances and the $^3$He isotopic anomaly of subdwarf B stars
}

\author{S. Geier \inst{1,2}
   \and U. Heber \inst{1}
   \and H. Edelmann \inst{1}
   \and L. Morales-Rueda \inst{3}
   \and D. Kilkenny \inst{4}
   \and D. O'Donoghue \inst{5,6}
   \and T. R. Marsh \inst{7}
   \and C. Copperwheat \inst{8,7}
   }

\offprints{S.\,Geier,\\ \email{sgeier@eso.org}}

\institute{Dr. Karl Remeis-Observatory \& ECAP, Astronomical Institute,
Friedrich-Alexander University Erlangen-Nuremberg, Sternwartstr. 7, D 96049 Bamberg, Germany 
   \and European Southern Observatory, Karl-Schwarzschild-Str. 2, 85748 Garching, Germany
   \and Department of Astrophysics, Faculty of Science, Radboud University Nijmegen, P.O. Box 9010, 6500 GL Nijmegen, NE
   \and Department of Physics, University of Western Cape, Private Bag X17, Bellville 7535, South Africa
   \and South African Astronomical Observatory, PO Box 9, 7935 Observatory, Cape Town, South Africa
   \and Southern African Large Telescope Foundation, PO Box 9, 7935 Observatory, Cape Town, South Africa
   \and Department of Physics, University of Warwick, Conventry CV4 7AL, UK
   \and Astrophysics Research Institute, Liverpool John Moores University, Twelve Quays House, Egerton Wharf, Birkenhead CH 41 1LD, UK}

\date{Received \ Accepted}

\abstract{Atmospheric parameters and helium abundances of 44 bright subdwarf B stars have been determined. More than half of our sample consists of newly discovered stars from the Edinburgh Cape survey. We showed that effective temperatures and surface gravities can be derived from high resolution echelle spectra with sufficient accuracy. Systematic uncertainties have been determined by comparing the parameters derived from the high resolution data with the ones derived from medium resolution spectra. Helium abundances have been measured with high accuracy. Besides the known correlation of helium abundance with temperature, two distinct sequences in helium abundance have been confirmed. Significant isotopic shifts of helium lines due to an enrichment in $^{3}$He have been found in the spectra of 8 sdBs. Most of these stars cluster in a small temperature range between $27\,000\,{\rm K}$ and $31\,000\,{\rm K}$ very similar to the known $^{3}$He-rich main sequence B stars, which also cluster in such a small strip, but at different temperatures. Both the helium sequences and the isotopic anomaly are discussed.

}

\maketitle

\section{Introduction \label{sec:intro}}

Subluminous B stars (sdBs) have similar colours and spectral characteristics
to main sequence stars of spectral type B, but are much less luminous.
Strong line broadening and the early confluence of the Balmer series is
caused by the high surface gravities ($\log\,g\simeq5.0-6.0$) of these
compact stars ($R_{\rm sdB}\simeq0.1-0.3\,R_{\rm \odot}$). SdBs are
considered to be core helium-burning stars with very thin hydrogen envelopes
and masses of about half a solar mass (Heber \cite{heber86}) located at the
extreme end of the horizontal branch (EHB).

The origin of these stars is unclear. A mass loss mechanism must manage to
remove all but a tiny fraction of the hydrogen envelope at about the same
time as the helium core has attained the mass ($\simeq0.5\,M_{\rm \odot}$)
required for the helium flash. The reason for this mass loss is still
unknown. Several single star scenarios invoke enhanced stellar winds or
interaction with the stellar environment (see Heber \cite{heber09} for a
review).

However, Mengel et al. (\cite{mengel76}) showed that the required strong
mass loss can occur in a close-binary system. The progenitor of the sdB star
has to fill its Roche lobe near the tip of the red-giant branch (RGB) to
lose most of its hydrogen-rich envelope. The merger of binary white dwarfs
was investigated by Webbink (\cite{webbink84}) who showed that an EHB star
can form when two helium core white dwarfs merge and the product is
sufficiently massive to ignite helium.

Maxted et al. (\cite{maxted01}) determined a very high fraction of radial
velocity variable sdB stars (see also Morales-Rueda et al. \cite{morales03};
Napiwotzki et al. \cite{napiwotzki04a}; Copperwheat et al.
\cite{copperwheat11}). Han et al. (\cite{han02,han03}) used binary
population synthesis models and studied the stable Roche lobe overflow
(RLOF) channel, the common envelope ejection (CE) channel, where the mass
transfer to the companion is dynamically unstable, and the He-WD merger
channel.

The formation of sdBs has also been related to the origin of the even more
enigmatic He-sdO/Bs (Ahmad \& Jeffery \cite{ahmad03}; Naslim et al.
\cite{naslim10}; Str\"oer et al. \cite{stroeer07}; Hirsch \& Heber
\cite{hirsch09}). The so-called late hot flasher scenario was proposed to
form these objects (Lanz et al. \cite{lanz04}; Miller-Bertolami et al.
\cite{miller08}).

At first glance, determining the helium abundance in sdB atmospheres seems
to be the best diagnostic tool to distinguish between the different
formation channels. While the merger of two He-WDs would form a pure
He-star, a wide range of helium abundances is predicted by the late hot
flasher scenario. Different mass-loss on the RGB, either triggered via
single-star or binary evolution, may also leave an imprint on the helium
abundance of the formed sdB.

Unfortunately, the primordial helium abundance of sdB stars is significantly
affected by processes in the hot and very dense atmospheres of these stars.
Sargent \& Searle (\cite{sargent66}) found sdB stars to be helium deficient
for the first time. Greenstein, Truran \& Cameron (\cite{greenstein67})
suggested that diffusion might cause the observed helium deficiency.
However, theoretical diffusion models yielded only little success (e.g.
Michaud, Vauclair \& Vauclair \cite{michaud83}), since the timescales for
the gravitational settling were predicted to be too short. The atmospheres
of sdBs should not only be depleted in helium, but essentially helium-free.
Several attempts have been made to model the atmospheres of sdBs by invoking
radiative levitation and mass loss caused by stellar winds to counteract the
gravitational settling (Bergeron et al. \cite{bergeron88}; Michaud et al.
\cite{michaud89}; Fontaine \& Chayer \cite{fontaine97}; Ohl et al.
\cite{ohl00}; Unglaub \& Bues \cite{unglaub01}).

Diffusion not only affects the elemental abundances, but can also lead to a
separation of different isotopes. When a heavier isotope is significantly
affected by gravitational settling, the lighter one appears to be enriched
in the atmosphere. An enrichment of $^{3}$He has initially been found in
main sequence B stars with subsolar helium abundance (Sargent \& Jugaku
\cite{sargent61}; Hartoog \& Cowley \cite{hartoogc79}) and explained by
diffusion processes (Vauclair et al. \cite{vauclair74}). Feige\,86 was the
first horizontal branch star showing this anomaly (Hartoog
\cite{hartoog79}). Eventually Heber (\cite{heber87}) detected strong line
shifts in the sdB star SB\,290 and the blue horizontal branch star
PHL\,25 indicating that basically the whole helium content of the atmosphere
consists of $^{3}$He. Later, Edelmann et al. (\cite{edelmann01}) and Heber
\& Edelmann (\cite{heber04}) found another three sdBs (Feige\,36,
BD\,+48\,2721, PG\,0133$+$114) where $^{3}$He is enriched in the atmosphere.

Finally, diffusion may change the atmospheres of hot subdwarfs in an even
more substantial way. Miller-Bertolami et al. (\cite{miller08}) argued that
due to diffusion processes He-sdOs will turn into hydrogen rich subdwarfs
before they evolve towards the white dwarf graveyard. The discovery of sdBs
with He abundances between the normal sdBs and the He-rich ones seems to be
consistent with this scenario (Ahmad et al. \cite{ahmad07}; {  Vennes et al. \cite{vennes07}}; 
Naslim et al. \cite{naslim11}; Naslim et al. \cite{naslim12}).

In this paper we determine the helium abundances and isotopic shifts caused
by enrichment of $^{3}$He of $44$ sdBs from high-resolution spectroscopy.
Previous papers of this series dealt with the rotational properties of sdB
binaries (Geier et al. \cite{geier10}, Paper I), the rotational properties
of single sdBs (Geier \& Heber \cite{geier12}, Paper II) and the metal
abundances of sdBs in the context of diffusion (Geier \cite{geier13}, Paper
III).

In Sect.~\ref{sec:obs} we give an overview of the observations taken with
different instruments. The determination of the atmospheric parameters and
helium abundances as well as an evaluation of the uncertainties {  are}
described in Sect.~\ref{sec:param}. In Sects.~\ref{sec:results_atmo},
\ref{sec:results_he} and \ref{sec:results_iso} we present our results
regarding the atmospheric parameters, helium abundances and isotopic shifts,
which are discussed further in Sect.~\ref{sec:discussion}. Finally, a
summary is given in Sect.~\ref{sec:summary}.

\section{Observations and Data Reduction \label{sec:obs}}

39 bright subdwarf B stars were observed with the FEROS spectrograph
($R=48\,000$, $3750-9200\,{\rm \AA}$) mounted at the ESO/MPG 2.2m telescope
in La Silla. The spectra were downloaded from the ESO science archive and
reduced with the FEROS-DRS pipeline under the ESO MIDAS context in optimum
extraction mode. The single spectra of all programme stars were RV-corrected
and co-added in order to achieve higher signal-to-noise.

Five stars were observed with the FOCES spectrograph ($R=30\,000$,
$3800-7000\,{\rm \AA}$) mounted at the CAHA 2.2m telescope {  and the 
spectra were also reduced} with the MIDAS package.

Medium resolution spectra of 12 stars were obtained with the ISIS
spectrograph ($R\simeq4000,\lambda=3440-5270\,{\rm \AA}$) mounted at the
WHT. Nine sdBs discovered in the course of the Edinburgh-Cape blue object
survey (Stobie et al. \cite{stobie97}; Kilkenny et al. \cite{kilkenny97};
O'Donoghue et al. \cite{odonoghue13}) have been observed with the grating
spectrograph and intensified {  Reticon Photon Counting System} on the
1.9m telescope of the SAAO ($R\simeq1300,\lambda=3300-5600\,{\rm \AA}$).
Spectra of five sdBs have been taken with the CAFOS spectrograph mounted at
the CAHA 2.2m telescope ($R\simeq1000,\lambda=3500-5800\,{\rm \AA}$).
Table~\ref{tab:data} provides a detailed overview {  of} the observed
sample and the data.

\begin{table*}[t!]
\caption{Target sample and data.}\label{tab:data}
\begin{center}
\begin{tabular}{llllllccl}
\hline
\noalign{\smallskip}
Object & other names & $m$\,[mag] & Instrument & no. spec & S/N & ISIS & SAAO-Reticon & CAFOS \\
\noalign{\smallskip}
\hline
\noalign{\smallskip}
BD$+$48\,2721         & & 10.7$^{\rm V}$ & FOCES   & 1 & 84  &  &  & \\
$[$CW83$]$\,0512$-$08 & & 11.3$^{\rm V}$ & FEROS   & 2 & 66 &  & &  \\
$[$CW83$]$\,1758$+$36 & PG\,1758$+$364 & 11.4$^{\rm V}$ & FOCES   & 1 & 41  &  &  & \\
EC\,00042$-$2737      & & 14.0$^{\rm V}$ & FEROS   & 2 & 21 &  & + & \\
EC\,01120$-$5259      & & 13.5$^{\rm V}$ & FEROS   & 2 & 45  &  & + & \\  
EC\,02542$-$3019      & & 12.8$^{\rm B}$ & FEROS   & 2 & 39  &  & + & \\ 
EC\,03263$-$6403      & & 13.2$^{\rm V}$ & FEROS   & 1 & 17  &  &   & \\
EC\,03408$-$1315      & & 13.6$^{\rm V}$ & FEROS   & 3 & 29  &  &  & \\
EC\,03470$-$5039$^{\rm ir}$      & & 13.6$^{\rm V}$ & FEROS   & 2 & 31  &   & + & \\ 
EC\,03591$-$3232      & CD\,$-$32\,1567 & 11.2$^{\rm V}$     & FEROS   & 2 & 73  &  &   & \\
EC\,05479$-$5818      & & 13.1$^{\rm V}$ & FEROS   & 3 & 47  &  & + & \\ 
EC\,10189$-$1502      & & 13.8$^{\rm V}$ & FEROS   & 2 & 35 &  & & \\
EC\,12234$-$2607      & & 13.8$^{\rm V}$ & FEROS   & 3 & 26  &  &   & \\ 
EC\,13047$-$3049      & & 12.8$^{\rm V}$ & FEROS   & 2 & 47  &  & + & \\
EC\,14248$-$2647      & & 12.0$^{\rm V}$ & FEROS   & 1 & 60  &  & + & \\  
EC\,14338$-$1445$^{\rm rv}$      & &  13.5$^{\rm V}$  & FEROS   & 3 & 37  &  & + & \\
EC\,20106$-$5248      & & 12.6$^{\rm V}$ & FEROS   & 4 & 60  &  & + & \\
EC\,20229$-$3716      & & 11.4$^{\rm V}$ & FEROS   & 3 & 69  &  &  & \\ 
EC\,21043$-$4017      & & 13.1$^{\rm V}$ & FEROS   & 2 & 37  &  &  & \\   
EC\,22081$-$1916      & & 12.9$^{\rm V}$ & FEROS   & 3 & 40  &  &   & \\  
Feige\,38             & PG\,1114$+$072 & 12.8$^{\rm B}$ & FEROS   & 5 & 89  & + &  & + \\
Feige\,65             & PG\,1233$+$426 & 11.8$^{\rm B}$ & FOCES   & 1 & 54  &  &  & \\
GD\,108               & PG\,0958$-$072 & 13.3$^{\rm B}$ & FEROS   & 3 & 61  &  &  & + \\
HE\,0447$-$3654       & & 14.6$^{\rm V}$ & FEROS   & 1 & 26  &  &   & \\ 
LB\,1516$^{\rm l,rv}$ & & 12.7$^{\rm B}$ & FEROS   & 2 & 35  &  &  & \\
PB\,5333              & & 12.5$^{\rm B}$ & FEROS   & 1 & 42 & + & & + \\
PG\,0342$+$026        & & 11.1$^{\rm B}$ & FEROS   & 4 & 106 &  &  & \\
PG\,0909$+$164$^{\rm s}$  & & 13.9$^{\rm B}$ & FEROS   & 2 & 33  & + & & \\
PG\,0909$+$276        & & 13.9$^{\rm B}$ & FEROS   & 2 & 52 & & & + \\
PG\,1303$+$097        & & 14.3$^{\rm B}$ & FEROS   & 3 & 31  & + &  & \\  
PG\,1505$+$074        & & 12.2$^{\rm B}$ & FEROS   & 3 & 102 & + & & + \\
PG\,1519$+$640$^{\rm rv}$ & & 12.1$^{\rm B}$ & FOCES   & 1 & 39  & + &  & \\ 
PG\,1616$+$144        & & 13.5$^{\rm B}$ & FEROS   & 1 & 24  & + & & \\
PG\,1653$+$131        & & 14.1$^{\rm B}$ & FEROS   & 3 & 40  & + &  & \\
PG\,1710$+$490        & & 12.1$^{\rm B}$ & FOCES   & 1 & 27  & + &  & \\
PG\,2151$+$100        & & 12.9$^{\rm B}$ & FEROS   & 3 & 39  &  &  & \\   
PG\,2205$+$023        & & 12.9$^{\rm B}$ & FEROS   & 2 & 19  & + &  & \\
PG\,2314$+$076$^{\rm rv}$ & & 13.9$^{\rm B}$ & FEROS   & 2 & 41  & + & & \\ 
PG\,2349$+$002        & & 12.0$^{\rm B}$ & FEROS   & 2 & 50  & + &  & \\
PHL\,44$^{\rm l}$     & EC\,21324$-$1346 & 13.0$^{\rm B}$ & FEROS   & 3 & 44  &  &  & \\
PHL\,334              & TON\,S 61  & 12.5$^{\rm B}$ & FEROS   & 3 & 48  &  &  & \\ 
                      & BPS\,CS\,23431$-$0044 &      &         &   &     &  &  & \\ 
PHL\,457$^{\rm l,rv}$ & GD\,1110 & 13.0$^{\rm B}$ & FEROS   & 2 & 50  &  &  & \\ 
PHL\,1548             & & 12.5$^{\rm B}$ & FEROS   & 3 & 42 & & & \\
SB\,815               & CD\,$-$35\,15910 & 10.6$^{\rm B}$ & FEROS   & 2 & 50  &  &  & \\
\noalign{\smallskip}
\hline
\end{tabular}
\tablefoot{$^{\rm s}$Pulsating subdwarf of V\,361\,Hya type (sdBV$_{\rm r}$). $^{\rm l}$Pulsating subdwarf of
 V\,1093\,Her type (sdBV$_{\rm s}$). $^{\rm rv}$Radial velocity variable star in a close binary (PHL\,457, LB\,1516, Edelmann et al. \cite{edelmann05}; PG\,1519$+$640, PG\,2314$+$076, Copperwheat et al. \cite{copperwheat11}). A maximum RV shift of $54.0\pm1.4\,{\rm km\,s^{-1}}$ has been detected between the three FEROS spectra of EC\,14338$-$1445 analysed in this work. $^{\rm ir}$ An excess in the infrared 2MASS colours has been reported by Copperwheat et al. (\cite{copperwheat11}), which may be due to a late main sequence companion.}
\end{center}
\end{table*}
 
\section{Atmospheric parameter determination and systematic effects \label{sec:param}}

\begin{table*}[t!]
\caption{Atmospheric parameters}\label{tab:param}
\begin{center}
\begin{tabular}{llllllll}
\hline
\noalign{\smallskip}
System & Instrument & $T_{\rm eff}$ & $\log{g}$ & $\log{y}$ \\
       & & [K] &  &  & \\ 
\noalign{\smallskip}
\hline
\noalign{\smallskip}
EC\,20106$-$5248        & FEROS   & 24500 & 5.25 & -2.77 & \\
BD$+$48\,2721           & FOCES   & 24800 & 5.38 & -2.23 & $^{3}$He \\
LB\,1516                & FEROS   & 25200 & 5.41 & -2.78 & \\
PG\,1653$+$131          & ISIS    & 25400 & 5.41 & -2.70 & \\
PG\,0342$+$026          & FEROS   & 26000 & 5.59 & -2.69 & \\
GD\,108                 & CAFOS   & 26100 & 5.58 & -3.46 & \\
Feige\,65               & FOCES   & 26200 & 5.31 & -2.75 & \\
PHL\,457                & FEROS   & 26500 & 5.38 & -2.54 & \\
PHL\,44                 & FEROS   & 26600 & 5.41 & -2.97 & \\
SB\,815                 & FEROS   & 27000 & 5.32 & -2.90 & \\
PG\,2205$+$023          & ISIS    & 27100 & 5.51 & $<-4.0$ & \\
PG\,2314$+$076          & ISIS    & 27200 & 5.65 & $<-4.0$ & \\
EC\,14338$-$1445        & FEROS   & 27700 & 5.54 & -2.82 & $^{3}$He\\
EC\,03591$-$3232        & FEROS   & 28000 & 5.55 & -2.03 & $^{3}$He\\
EC\,12234$-$2607        & FEROS   & 28000 & 5.58 & -1.58 & $^{3}$He \\
PG\,2349$+$002          & ISIS    & 28000 & 5.73 & -3.45 & \\
EC\,01120$-$5259        & FEROS   & 28900 & 5.41 & -2.54 & \\
EC\,03263$-$6403        & FEROS   & 29300 & 5.48 & -2.51 & $^{3}$He \\
PG\,1303$+$097          & ISIS    & 29800 & 5.83 & -2.17 & \\
PG\,1519$+$640          & ISIS    & 30300 & 5.67 & -2.37 & $^{3}$He \\
EC\,03470$-$5039        & FEROS   & 30500 & 5.61 & $<-4.0$ & \\
PG\,1710$+$490          & ISIS    & 30600 & 5.66 & -2.43  & $^{3}$He \\
Feige\,38               & ISIS    & 30600 & 5.83 & -2.37 & $^{3}$He \\
HE\,0447$-$3656         & FEROS   & 30700 & 5.57 & $<-3.0$ & \\
EC\,22081$-$1916        & FEROS   & 31100 & 4.77 & -1.97 & \\
EC\,14248$-$2647        & FEROS   & 31400 & 5.56 & -1.64 &  \\
EC\,02542$-$3019        & FEROS   & 31900 & 5.68 & -1.89 &  \\
EC\,21043$-$4017        & FEROS   & 32400 & 5.63 & -1.58 &  \\
EC\,20229$-$3716        & FEROS   & 32500 & 5.00 & -1.75 & \\
PG\,2151$+$100          & FEROS   & 32700 & 5.59 & $<-3.0$ & \\
EC\,05479$-$5818        & FEROS   & 33000 & 5.93 & -1.66 & \\
EC\,13047$-$3049        & FEROS   & 34700 & 5.35 & -2.57 & \\
$[$CW83$]$\,1758$+$36   & FOCES   & 34600 & 5.79 & -1.51 & \\
PHL\,334                & FEROS   & 34800 & 5.84 & -1.42 & \\
PG\,0909$+$164          & ISIS    & 35300 & 5.33 & -2.76 & \\
PG\,0909$+$276          & CAFOS   & 35500 & 6.09 & -1.00 & \\
EC\,03408$-$1315        & FEROS   & 35700 & 5.85 & -1.61 & \\
PG\,1505$+$074          & ISIS    & 37100 & 5.39 & -2.69 & \\
PG\,1616$+$144          & ISIS    & 37300 & 5.95 & -1.26 & \\
PHL\,1548               & FEROS   & 37400 & 5.79 & -1.55 & \\
EC\,00042$-$2737        & FEROS   & 37500 & 5.94 & -1.62 & \\
EC\,10189$-$1502        & FEROS   & 37900 & 5.43 & -2.28 & \\
$[$CW83$]$\,0512$-$08   & FEROS   & 38400 & 5.77 & -0.73 & \\
PB\,5333                & ISIS    & 40600 & 5.96 & -2.62 & \\
\noalign{\smallskip}
\hline
\end{tabular}
\end{center}
\end{table*}

\begin{figure}[t!]
\begin{center}
	\resizebox{\hsize}{!}{\includegraphics{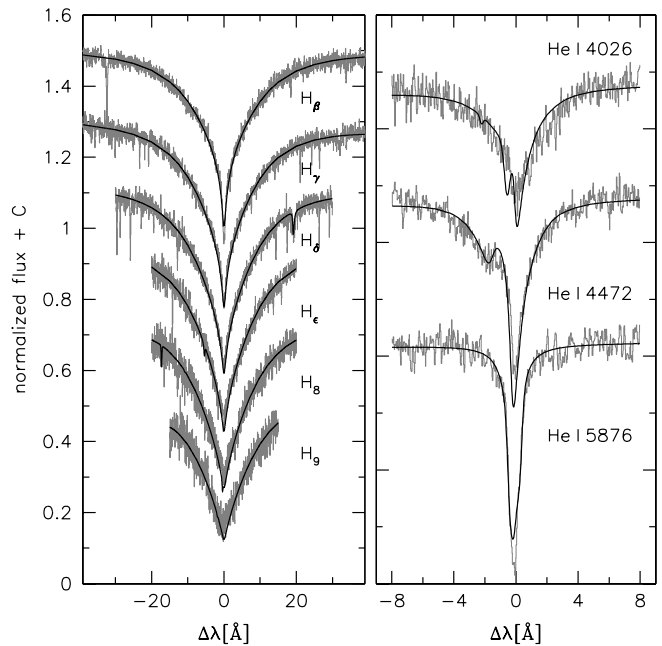}}
	\caption{Example fit of LTE models to the Balmer and selected He\,{\sc i} lines in the FEROS spectrum of EC\,03591$-$3232.}
	\label{fig:example}
\end{center}
\end{figure}

\begin{figure}[t!]
\begin{center}
	\resizebox{\hsize}{!}{\includegraphics{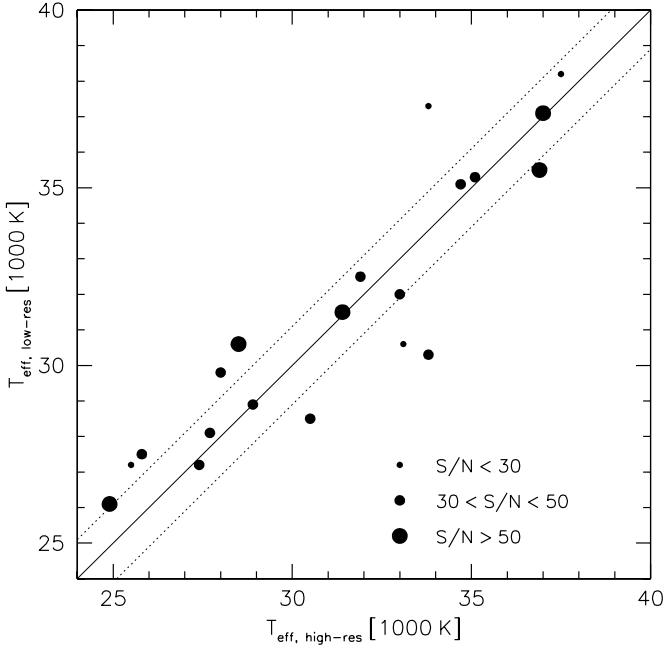}}
	\caption{Effective temperatures derived from high-resolution spectra plotted against effective temperature derived from medium resolution spectra (ISIS, SAAO-Reticon, CAFOS). The dotted lines mark the average deviation between the two datasets. The size of the points scales with the data quality of the
	high-resolution spectra.}
	\label{fig:teffvsteff_lowres}
\end{center}
\end{figure}

\begin{figure}[t!]
\begin{center}
	\resizebox{\hsize}{!}{\includegraphics{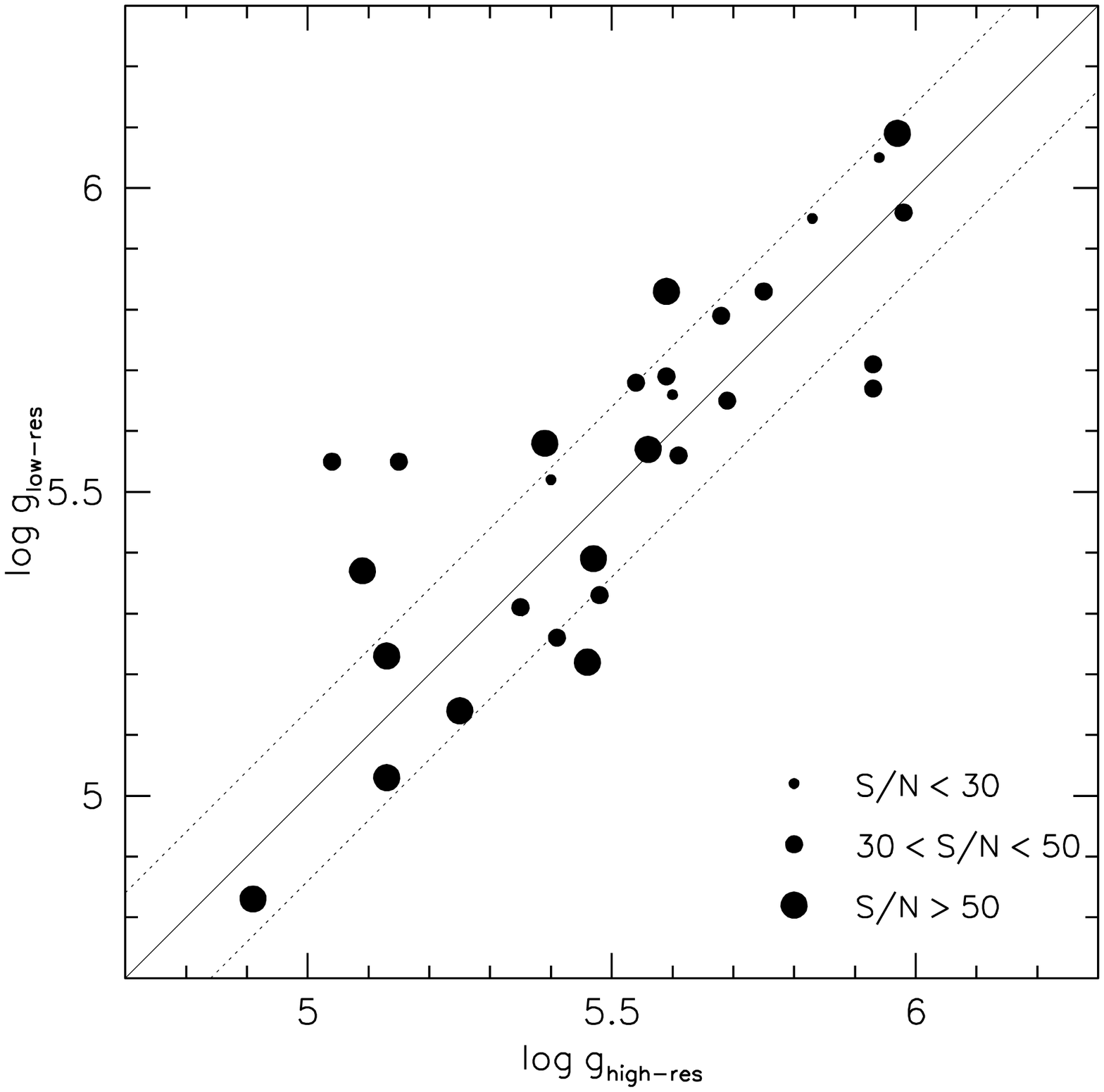}}
	\caption{Surface gravity derived from high-resolution spectra plotted against surface gravity derived from medium-resolution spectra (see Fig.~\ref{fig:teffvsteff_lowres}).}
	\label{fig:loggvslogg_lowres}
\end{center}
\end{figure}

\subsection{High-resolution echelle spectra - known issues}\label{sec:issues}

Most spectroscopic studies of sdB stars {  have} relied on low and
medium-resolution data ($R\simeq500-4000$). Fitting model atmospheres to
high S/N data allows us to determine $T_{\rm eff}$ and $\log{g}$ with formal
uncertainties lower than the systematic effects between different model
grids (Green et al. \cite{green08}; Heber et al. \cite{heber00}). This
standard technique yielded key results on which our current understanding of
sdB formation and evolution is founded (e.g. Heber \cite{heber86}; Heber \&
Langhans \cite{heber86b}; Saffer et al. \cite{saffer94}; Maxted et al.
\cite{maxted01}; Edelmann et al. \cite{edelmann03}; Green et al.
\cite{green08}; \O stensen et al. \cite{oestensen10}; Copperwheat et al.
\cite{copperwheat11}; Vennes et al. \cite{vennes11}; Geier et al.
\cite{geier11b}; N\'emeth et al. \cite{nemeth12}).

On the other hand, high resolution echelle spectrographs ($R>10\,000$) are
widely used in astronomy today. However, data reduction and analysis of
echelle spectra is difficult. Even if sophisticated data reduction pipelines
are available, issues {  such as} fringing, extraction errors,
insufficient order merging {  and} normalisation remain. Due to the low
luminosity of sdBs, echelle spectroscopy of these objects is rather
challenging, {  as only a few of them have magnitudes brighter than 10th}.
Furthermore, sdBs are hot stars and the most important lines for their
analysis are found in the bluest parts of optical spectra; {  
unfortunately, echelle spectrographs are often not very sensitive in the
blue, particularly where fibre-feeds are used.}

The spectra of sdBs are dominated by strong and broad hydrogen Balmer lines,
which are the key {  to deriving} their atmospheric parameters. Since
these lines are usually broader than the single echelle orders, merging
errors can severely affect the parameter determination {  and} suspicious
features can be hard to spot, particularly in low S/N data. However, Lisker
et al. (\cite{lisker05}) successfully analysed high resolution spectra of
sdBs observed in the course of the ESO-SPY project (Napiwotzki et al.
\cite{napiwotzki03}). In this case, the UVES spectrograph at the ESO-VLT
{  was} used, which is very sensitive in the blue wavelength range.
Furthermore, rectification of the data was done by dividing {  by the}
featureless spectra of DC-type white dwarfs. However, the {  most of the} known DC white
dwarfs are too faint to be observed with 2m-class telescopes and it is
therefore not clear, {  in general}, whether reliable atmospheric
parameters of sdBs can be derived from medium S/N ($30-50$), high resolution
echelle spectra.

\subsection{Determination of atmospheric parameters}

Atmospheric parameters and helium abundances of our sample have been
determined by fitting model spectra to the hydrogen Balmer and helium lines
of the high-resolution spectra using the SPAS routine developed by H. Hirsch
(see Fig.~\ref{fig:example}). The method is described in Copperwheat et al.
(\cite{copperwheat11}). {  To derive the atmospheric parameters we used LTE model 
with solar metallicity for stars with effective temperatures between $24000\,{\rm K}$ and $30000\,{\rm K}$. 
The hottest stars with $T_{\rm eff}>30000\,{\rm K}$ have been fitted with LTE models 
with supersolar metallicity (see Copperwheat et al. \cite{copperwheat11}).} 

Due to the high resolution of the spectra the formal
errors derived with a bootstrapping algorithm are much smaller than the
typical systematic offsets between different model grids ($\Delta T_{\rm
eff}\simeq500\,{\rm K}$, $\Delta \log{g}\simeq0.05$). However, these formal
uncertainities must not be adopted as error estimates, because systematic
shifts are the dominant error source in this case.

\subsection{Comparison with medium resolution data}

Due to the caveats discussed in Sect.~\ref{sec:issues}, the parameter
determination from the high-resolution spectra needs to be checked and
systematic effects have to be quantified properly. We used medium-resolution
spectra obtained with the ISIS, the SAAO-Reticon and the CAFOS spectrograph
and analysed them in the same way as the high-resolution data.

The ISIS and CAFOS spectra have a very high {  S/N ($>100$)} and cover the
blue spectral range down to the Balmer jump. These spectra are perfectly
suited for the determination of atmospheric parameters. Again, formal errors
are much smaller than the typical systematic offsets between different model
grids, which are adopted as error estimates in this case. The SAAO-Reticon
spectra also cover the higher Balmer lines, but their quality is
inhomogeneous. Although the formal fitting errors {  can be large --
ranging from $700\,{\rm K}$ to $2000\,{\rm K}$ in $T_{\rm eff}$ and from
$0.12$ to $0.28$ in $\log{g}$ --} they provide a valuable consistency check
for the results derived from the high-resolution data.

Figs.~\ref{fig:teffvsteff_lowres} and \ref{fig:loggvslogg_lowres} show the comparison between the high-resolution and the medium-resolution parameters. In general, the consistency between the parameters derived from high- and medium-resolution data is reasonable.  

\begin{figure}[t!]
\begin{center}
	\resizebox{\hsize}{!}{\includegraphics{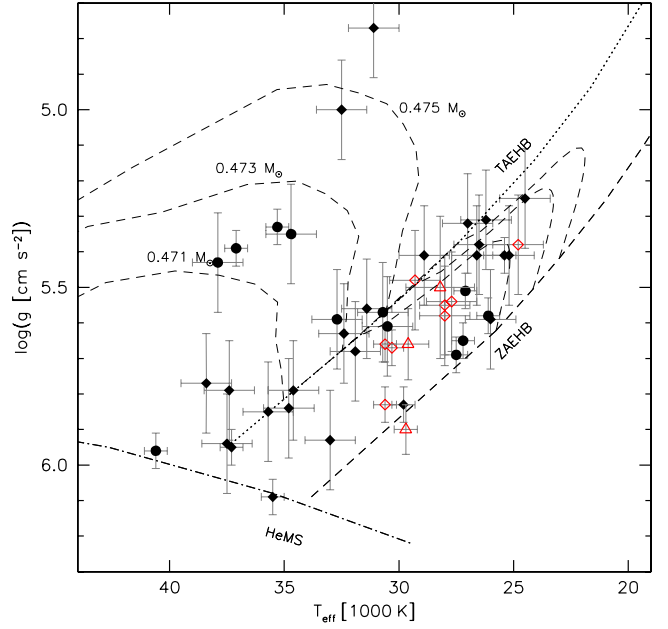}}
	\caption{$T_{\rm eff}-\log{g}$-diagram for the entire sample under study. 
	 The helium main sequence (HeMS) and the EHB band (limited by the zero-age 
         EHB, ZAEHB, and the terminal-age EHB, TAEHB) are superimposed with EHB evolutionary tracks for solar metallicity taken from Dorman et al. (\cite{dorman93}) labelled with their masses. Open symbols mark objects where isotopic shifts due to an enrichment of $^{3}$He were detected, filled symbols objects with atmospheres dominated by $^{4}$He. The diamonds mark stars belonging to the upper helium sequence, the circles stars belonging to the lower sequence (see Fig.~\ref{fig:nhevsteff}). The triangles mark the three sdBs with enriched $^{3}$He from literature (Heber et al. \cite{heber84}; Edelmann et al. \cite{edelmann99}; Morales-Rueda et al. \cite{morales03}).}
	\label{fig:tefflogg}
\end{center}
\end{figure}

\begin{figure*}[t!]
\begin{center}
	\resizebox{\hsize}{!}{\includegraphics{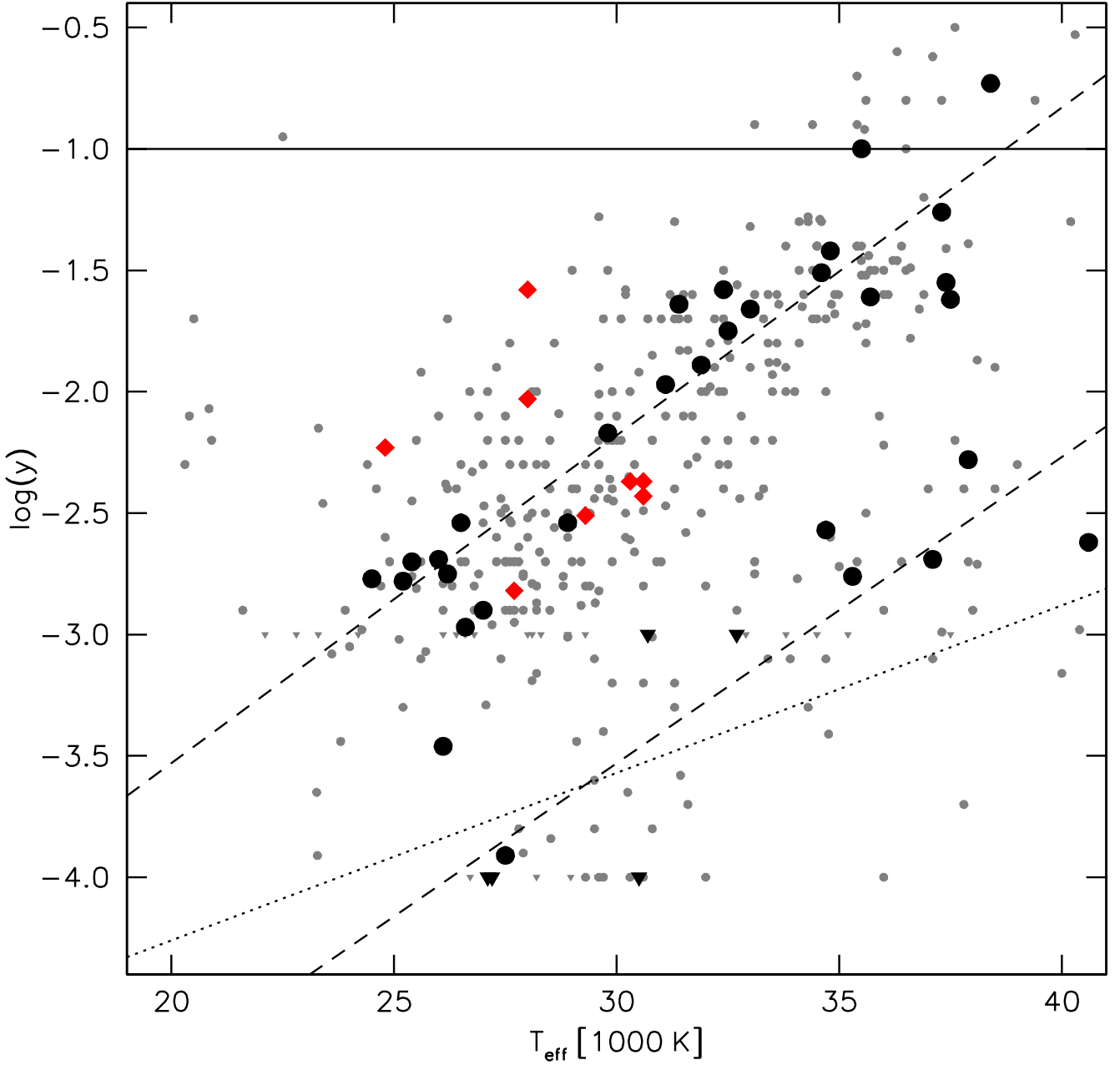}}
	\caption{Helium abundance $\log{y}$ plotted against effective temperature. The filled symbols mark the results from our study. Filled diamonds mark objects where isotopic shifts due to an enrichment of $^{3}$He were detected, filled circles objects with atmospheres dominated by $^{4}$He. Upper limits are marked with triangles. The solid horizontal line is drawn at solar helium abundance. The two dashed lines are regression lines for the two distinct helium sequences taken from Edelmann et al. (\cite{edelmann03}). The dotted regression line for the lower sequence is taken from N\'emeth et al. (\cite{nemeth12}). Measurements taken from literature are plotted as grey symbols.}
	\label{fig:nhevsteff}
\end{center}
\end{figure*}

In order to quantify the uncertainties in the parameter determination, the
averages of the shifts with respect to the parameters derived from
medium-resolution spectra {  have been calculated; these are} $\Delta
T_{\rm eff}\simeq1100\,{\rm K}$ and $\Delta \log{g}\simeq0.12$. The high
resolution spectra of PG\,1616+144, PG\,1710+490 and PG\,2205+023 have S/N
values below $30$ and hence show large deviations especially in $T_{\rm
eff}$ (up to $3500\,{\rm K}$) {  and these} stars have been excluded
before calculating the average. Although not perfect, {  the above}
uncertainties are consistent with values found in the literature (see { 
Appendix A}).

\section{Atmospheric parameters}\label{sec:results_atmo}

The atmospheric parameters of all 44 programme stars are shown in
Table~\ref{tab:param}. The final parameters are derived either from high S/N
ISIS and CAFOS spectra, if available, or from high-resolution spectra
obtained with FEROS and FOCES. The $T_{\rm eff}-\log{g}$-diagram of the
whole sample under study is shown in Fig.~\ref{fig:tefflogg}. All stars are
concentrated on or above the EHB, fully consistent with theory{ , and we}
therefore conclude that the atmospheric parameter determination is of
sufficient quality {  (see also the comparison with independent determinations 
from the literature given in Appendix Table A.1)}.

\begin{figure}[t!]
\begin{center}
	\resizebox{\hsize}{!}{\includegraphics{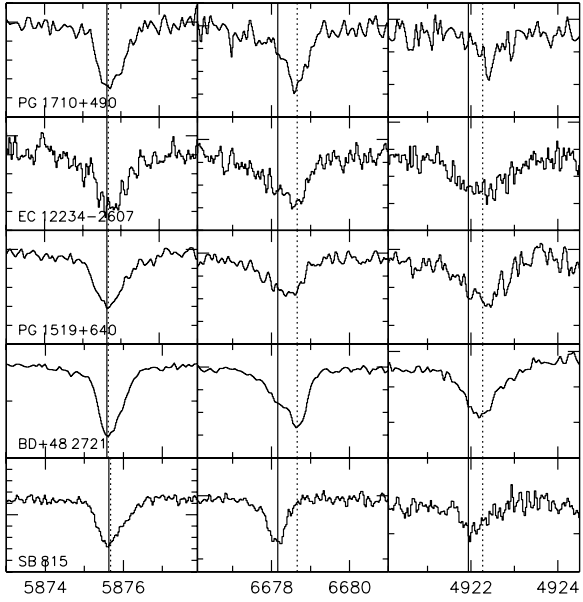}}
	\caption{Helium lines of sdB stars. The rest wavelengths of the $^{4}$He (solid) and $^{3}$He lines (dotted) are  plotted as vertical lines. The rest wavelengths of the $^{3}$He lines have been taken from Hartoog \& Cowley (\cite{hartoogc79}). The stars are highly enriched in $^{3}$He, but also show a component of $^{4}$He. For comparison, the He lines of SB\,815 are plotted in the lowest panel. This star does not show line shifts due to enrichment of $^{3}$He.}
        \label{fig:heshift1}
\end{center}
\end{figure}

\section{Helium abundances}\label{sec:results_he}

High resolution spectra are very well suited {  for measuring} accurate
elemental abundances of sdBs, because the rather weak lines of helium and
metals are fully resolved. We therefore used the FEROS and FOCES spectra to
determine the helium abundances of our programme stars. The formal
uncertainties are very small ($\Delta\log{y}=0.01-0.07$) and comparable to
the deviations measured when analysing several single spectra of one object.
Taking systematic effects into account the helium abundances should
therefore be accurate to $\simeq0.1\,{\rm dex}$. The results are given in
Table~\ref{tab:param}.

Fig.~\ref{fig:nhevsteff} shows the helium abundances of our sample plotted
against effective temperature. All but two of our programme stars have
{  the} subsolar helium abundances typical {  of} sdB stars. Edelmann et al.
(\cite{edelmann03}) reported a correlation of helium abundance with
temperature, which was subsequently confirmed by Lisker et al.
(\cite{lisker05}) and N\'emeth et al. (\cite{nemeth12}). This correlation
can be clearly seen in our sample as well.

Edelmann et al. (\cite{edelmann03}) also reported the discovery of two
distinct sequences showing a similar correlation with temperature. The 'low
sequence' is offset by about $2\,{\rm dex}$ from the 'high sequence'. Lisker
et al. (\cite{lisker05}) and Geier et al. (\cite{geier11b}) could not
confirm this finding, but the sample size of these studies was smaller than
{  that} of Edelmann et al. (\cite{edelmann03}). N\'emeth et al.
(\cite{nemeth12}) studied a sample of bright hot subdwarfs spanning the
whole temperature and helium abundance range from sdBs to He-sdOs and found
indications for the two distinct sequences {  although,} in their sample,
the lower sequence appears to be less steep than reported by Edelmann et al.
(\cite{edelmann03}).

\begin{figure}[t!]
\begin{center}
	\resizebox{\hsize}{!}{\includegraphics{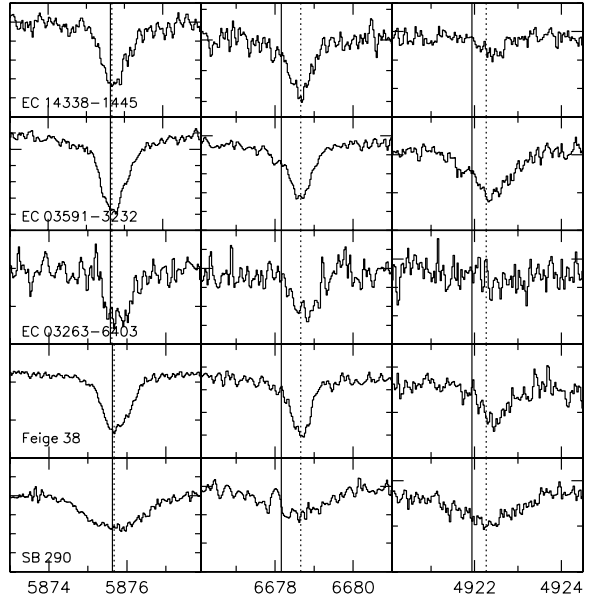}}
	\caption{{  Comments are as for Fig.~\ref{fig:heshift1} except
	that in these stars,} the helium lines are entirely
	shifted to the $^{3}$He rest wavelength and no traces of $^{4}$He
	are visible. For comparison, the He lines of SB\,290 are plotted in
	the lowest panel. This star is the prototype of sdBs enriched in
	$^{3}$He (Heber \cite{heber87}) {  but note that the lines show 
        significant rotational broadening} (Geier et al. \cite{geier13b}).}
        \label{fig:heshift2}
\end{center}
\end{figure}

The two distinct sequences are also visible in our data {  and combining
these} with the results of other studies (Saffer et al. \cite{saffer94};
Maxted et al. \cite{maxted01}; Edelmann et al. \cite{edelmann03};
Morales-Rueda et al. \cite{morales03}; Lisker et al. \cite{lisker05}; \O
stensen et al. \cite{oestensen10}; Geier et al. \cite{geier11b}; N\'emeth et
al. \cite{nemeth12}) the underlying pattern becomes more apparent. In
Fig.~\ref{fig:nhevsteff}, {  all of these} results are overplotted with the
two regression lines calculated by Edelmann et al. (\cite{edelmann03}) 
as well as the regression line for the lower sequence
calculated by N\'emeth et al. (\cite{nemeth12}) based on their results
spanning a larger range in effective temperatures and helium abundances. The
two lines by Edelmann et al. (\cite{edelmann03}) match very well with the
sequences seen in our sample, while the line by N\'emeth et al.
(\cite{nemeth12}) is slightly different. However, as has been correctly
pointed out by N\'emeth et al. (\cite{nemeth12}), those lines are only very
crude tentative models, which {  certainly} do not reflect the real
complexity of the underlying data.

Defining a dividing line between the two helium sequences by eye, which
follows the relation $\log{y}=0.127\,T_{\rm eff}/1000\,{\rm K}-6.718$, the
numbers of stars belonging to the different sequences can be counted. From
our sample of 44 stars, 31 {  of them} ($70\%$) are associated with the
upper sequence while 13 ($30\%$) belong to the lower. The respective
fractions of the full sample of $383$ sdBs are $82\%$ and $18\%$, {  but}
the full sample is likely to be biased against low helium abundances because
most analyses are based on low- and medium-resolution spectra where weak He
lines are {  often} not detectable. We therefore regard the respective
fractions derived from our sample to be more reliable.

\section{The $^{3}$He isotopic anomaly}\label{sec:results_iso}

The high-resolution spectra are {  also} perfectly suited to search for
small shifts in the rest wavelengths of the helium lines due to the
enrichment of $^{3}$He. {  Wavelength shifts can be caused by different
effects (for example, the presence of magnetic fields or pressure shifts in
high density environments), but} the helium line shifts caused by the
enrichment of $^{3}$He can be modelled quite accurately and show a typical
pattern. While some lines like He\,{\sc i}\,5876 are only shifted by
$0.04\,{\rm \AA}$ towards redder wavelengths, the shifts of He\,{\sc
i}\,4922 and He\,{\sc i}\,6678 are significant ($0.33$ and $0.50\,{\rm \AA}$
respectively; Fred et al. \cite{fred51}, given in Hartoog \& Cowley
\cite{hartoogc79}). Displacements of this order can be easily detected in
high resolution spectra.

All {  of the stars in} our sample have been examined {  and} in 8
cases, isotopic shifts due to the presence of $^{3}$He are clearly visible
(Figs.~\ref{fig:heshift1}, \ref{fig:heshift2}). One of these (BD+48\,2721)
has already been discovered by Edelmann et al. (\cite{edelmann01}).

\begin{figure}[t!]
\begin{center}
	\resizebox{\hsize}{!}{\includegraphics{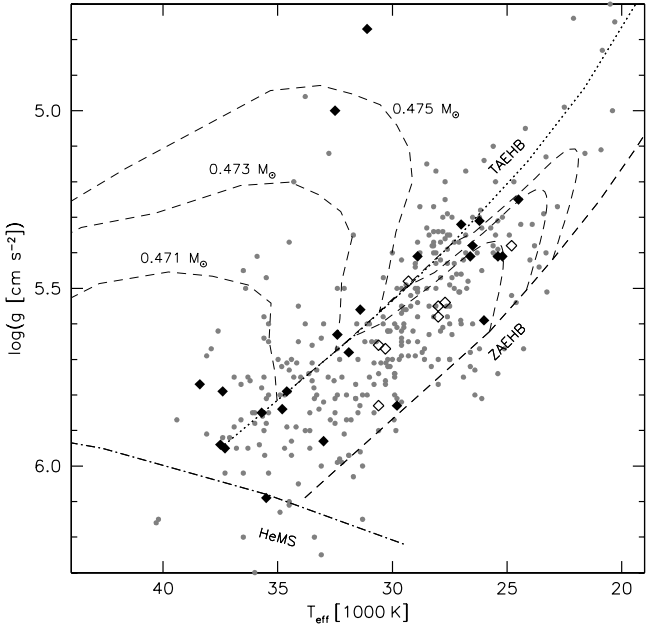}}
	\caption{$T_{\rm eff}-\log{g}$-diagram for the sdBs associated with the upper helium sequence (see Fig.~\ref{fig:tefflogg}).}
	\label{fig:tefflogg_high}
\end{center}
\end{figure}

\section{Discussion}\label{sec:discussion}

\subsection{Helium sequences}

The reasons for the correlation of the helium abundance with temperature and
the bimodal structure in the $T_{\rm eff}-\log{y}$ diagram {  are unknown,
although several suggestions have been made and are discussed briefly below}

Photospheric convection has been proposed {  as the cause of} the relative
enrichment of helium in sdB atmospheres towards higher temperature.
Greenstein \& Sargent (\cite{greenstein74}) suggested that a
He$^{+}$/He$^{2+}$ convection zone could transport helium from deeper layers
into the photosphere of subdwarfs hotter than $30\,000\,{\rm K}$ (but see
also Groth et al. \cite{groth85}). However, Michaud et al.
(\cite{michaud11}) calculated complete stellar evolution models, including
the effects of atomic diffusion and radiative acceleration, to study the
abundance anomalies observed on the hot end of the HB. Their models{ ,
which assume} extra mixing but no stellar-wind mass loss, are in general
agreement with {  the} observed metal abundances of sdB stars (Geier
\cite{geier13}) as well as the helium abundances for $T_{\rm
eff}>25\,000\,{\rm K}$. Furthermore, they show that diffusion effects reach
far deeper than the stellar atmospheres themselves and should also be
dominant in the He$^{+}$/He$^{2+}$ convection zone (Moehler, Michaud priv.
comm.).

According to {  the Michaud et al. (\cite{michaud11}) models,} the
observed scatter in helium abundance is caused by different HB ages and
differences in the initial metallicity of the progenitor populations, { 
but these models predict neither} the observed correlation with temperature
nor the two helium sequences. If the helium abundance {  does} depend on
the age of the sdB, one would expect to see a continuous distribution and
not a concentration {  in} distinct sequences.

\begin{figure}[t!]
\begin{center}
	\resizebox{\hsize}{!}{\includegraphics{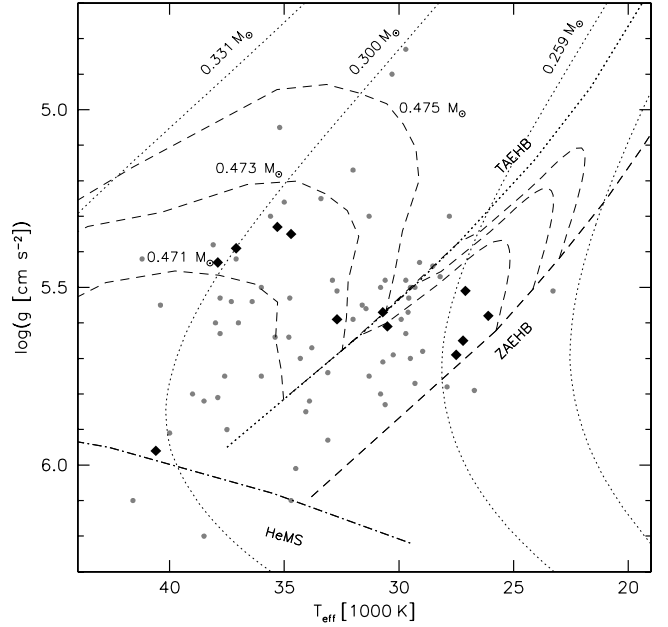}}
	
        \caption{$T_{\rm eff}-\log{g}$-diagram for the sdBs associated with
         the lower helium sequence. The figure is similar to
         Fig.~\ref{fig:tefflogg} {  with additional evolutionary tracks for
         post-RGB objects plotted} as dotted lines (Driebe et al.
         \cite{driebe98}).}

	\label{fig:tefflogg_low}
\end{center}
\end{figure}

Aznar Cuadrado \& Jeffery (\cite{aznar02}) argued that, due to tidal
effects, sdBs residing in short-period binaries {  might} have higher
photospheric helium content than long-period systems or single stars.
Edelmann et al. (\cite{edelmann03}) suggested that this effect {  could}
be responsible for the two helium sequences {  -- and this can be directly
tested} with the available data. Our {  current sample of bright sdBs
consists mostly} of stars {  in which} no radial velocity shifts have been
detected {  and so} these sdBs are not in close binaries with stellar mass
companions. {  On the other hand,} the target sample for the MUCHFUSS
project (Geier et al. \cite{geier11b}), which aims at finding sdBs with
massive compact companions in close orbits, consists {  only} of RV
variable stars. From the 51 sdBs drawn from {  the MUCHFUSS} sample, 15
belong to the lower helium sequence, {  so} the fraction of these stars is
almost exactly the same as in the sample presented here. The close
binary hypothesis can therefore be excluded as an explanation of the helium
sequences.

O'Toole (\cite{otoole08}) proposed another {  possibility: whilst} the
stars belonging to the upper sequence might be core helium-burning
(post-)EHB stars evolving in the way modelled by, {  for example,} Dorman
et al. (\cite{dorman93}), the sdBs forming the lower sequence {  could} be
post-RGB objects without helium burning in their cores. {  The latter}
objects are direct progenitors of He-WDs with masses ranging from
$\sim0.2\,M_{\rm \odot}$ to $0.33\,M_{\rm \odot}$ and are crossing the EHB
on cooling tracks (e.g. Driebe et al. \cite{driebe98}). The only sdB for
which the post-RGB nature could be proven unambiguously (using a
trigonometric parallax) is HD\,188112 (Heber et al. \cite{heber03}) { 
which} also has the lowest helium abundance ($\log{y}=-5$) ever measured for
an sdB. Two candidates of similar low mass have been discovered by Vennes at
al. (\cite{vennes11}) and Silvotti et al. (\cite{silvotti12}) {  but, even 
though these sdBs also show no detectable helium lines, their
post-RGB nature is less well established} than in the case of HD\,188112.
Observational evidence for particularly low helium abundance in post-RGB
objects therefore remains weak.

Another way to probe the post-RGB scenario is the comparison of both helium
populations in the $T_{\rm eff}$-$\log{g}$ diagram (O'Toole
\cite{otoole08}). In Fig.~\ref{fig:tefflogg_high}, the stars associated with
the upper sequence are plotted. Most sdBs are situated within or close to
the EHB band {  but about $10\%$ are above the EHB -- consistent with the
standard scenario in which} most stars are core helium-burning and therefore
residing on the EHB and only {  a minority} are more evolved shell
helium-burning objects (Dorman et al. \cite{dorman93}).

The distribution of the helium-poor sdBs {  is} different (see
Fig.~\ref{fig:tefflogg_low}) {  in that} the number density of stars on
and above the EHB is similar. In the classical picture, {  sdBs evolve
from the EHB but such an even distribution is not expected because} the
evolutionary time on the EHB is significantly longer than {  that for}
post-EHB evolution ($\sim10$ times as long),.

Comparing the distribution of {  the helium-poor} sdBs with post-RGB
tracks calculated by Driebe et al. (\cite{driebe98}), one would expect these
objects to be the progenitors of He-WDs with masses between $0.25\,{\rm
M_{\odot}}$ and $0.3\,{\rm M_{\odot}}$. The evolutionary timescales on these
tracks scale strongly with mass; while the evolution of an $0.25\,{\rm
M_{\odot}}$ object takes of the order of $100\,{\rm Myr}$, it shortens to
only a few million years if the object is more massive ($0.3\,{\rm
M_{\odot}}$). Accordingly, one would expect a higher density of objects with
low masses in the $T_{\rm eff}$-$\log{g}$ diagram {  (if the formation
rate of post-RGB objects does not depend on the mass) but this is not the
case -- most} objects are evenly distributed between the $\sim0.25\,{\rm
M_{\odot}}$ and $0.3\,{\rm M_{\odot}}$ evolutionary tracks (see
Fig.~\ref{fig:tefflogg_low}).

Being aware of this problem, O'Toole (\cite{otoole08}) argued that many more
He-WDs with masses close to $0.3\,{\rm M_{\odot}}$ are formed than objects
with even lower masses (e.g. Liebert et al. \cite{liebert05}; De Gennaro et
al. \cite{degennaro08}) {  which could} compensate for the difference in
evolutionary timescales. Given that the $18-29\%$ of the sdB population
belonging to the helium poor sequence are all direct He-WD progenitors with
an evolution about 100 times shorter than the lifetime on the EHB
($\simeq100\,{\rm Myr}$), the formation rate of these objects { 
(neglecting selection effects)} would be more than $\sim 20$ times higher
than the formation rate of core helium-burning sdBs. {  Even so}, the
distribution of objects seen in Fig.~\ref{fig:tefflogg_low} is still hard to
explain with the post-RGB scenario. The crowding near the EHB remains
especially suspicious, because evolution should be more or less uniform
along the tracks given by Driebe et al. (\cite{driebe98}). The lack of
objects below the EHB poses a particular problem in this respect. In
contrast, a significant number of He-WD progenitors with masses between
$0.16\,M_{\rm \odot}$ and $0.4\,M_{\rm \odot}$ have been discovered at
temperatures below $\simeq25\,000\,{\rm K}$ and well below the HB (see Kilic
et al. \cite{kilic12} and references therein). We therefore conclude that
{  post-RGB evolution}, despite being a very interesting option, is not able
to explain all {  of the} observations in a conclusive way.

\subsection{Isotopic anomaly}

The enrichment of $^{3}$He in {  around} 18\% of our programme stars { 
also} remains unexplained. Fig.~\ref{fig:tefflogg} shows the distribution of
these stars in the $T_{\rm eff}$-$\log{g}$-diagram, including the three sdBs
with isotopic shifts taken from the literature. It can be clearly seen that
they cluster in a narrow temperature range between $27\,000\,{\rm K}$ and
$31\,000\,{\rm K}$, with BD+48\,2721 ($T_{\rm eff}=24\,800\,{\rm K}$) being
the only exception {  and -- given the uncertainties -- this
$^{3}$He-strip could be pure}. The fact that all $^{3}$He enriched sdBs
belong to the upper helium sequence (Fig.~\ref{fig:nhevsteff}) {  might}
be a selection effect, because the diagnostic helium lines are too weak to
measure the isotopic shifts {  in the helium-poor stars}.

Most {  $^{3}$He sdB} stars show clear shifts of the He\,{\sc i} line at
$6678\,{\rm \AA}$, indicating that almost all {  the} helium in the
atmosphere is $^{3}$He. BD+48\,2721, EC\,12234$-$2607 and PG\,1519+640 show
strong lines of $^{3}$He blended with weak components of $^{4}$He (see
Fig.~\ref{fig:heshift1}) {  and} these three stars cover the whole $^{3}$He
temperature strip. The isotope ratio is therefore not correlated {  with} 
effective temperature.

Michaud et al. (\cite{michaud11}) predict a mild enrichment of $^{3}$He but,
due to gravitational settling of the heavier isotope, this should be the
case in all sdBs. Hartoog \& Cowley (\cite{hartoogc79}) studied the
enrichment of $^{3}$He in main sequence B stars and discovered a pattern
strikingly similar to our results{ :} stars enriched in $^{3}$He were
found at effective temperatures between $18\,000\,{\rm K}$ and
$21\,000\,{\rm K}${  ;} stars with lower temperatures down to
$\sim13\,000\,{\rm K}$ show slight underabundances in helium with respect to
the Sun, while the hotter stars up to $\sim32\,000\,{\rm K}$ are slighty
overabundant in helium.  The two known BHB stars with detected $^{3}$He
isotopic shifts (Feige\,86, $T_{\rm eff}=16\,400\,{\rm K}$, Bonifacio et al.
\cite{bonifacio95}; PHL\,25, $T_{\rm eff}=19\,500\,{\rm K}$, Heber \&
Langhans \cite{heber86b}) have temperatures close to the strip detected by
Hartoog \& Cowley (\cite{hartoog79}).

In Figs.~\ref{fig:tefflogg} and \ref{fig:nhevsteff} a similar pattern can be
seen. The sdBs enriched in $^{3}$He occupy a small strip in $T_{\rm eff}$,
{  within which} the helium abundance {  decreases} towards lower
temperatures and {  increases} towards higher temperatures.

Hartoog \& Cowley (\cite{hartoogc79}) argued that diffusion is responsible
for this effect. At low temperatures, the radiation pressure is not strong
enough to support helium in the atmosphere. As soon as the temperature
reaches a certain threshold value, the less massive $^{3}$He can be
supported, but not the more abundant $^{4}$He, {  which} leads to an
enrichment of $^{3}$He in the atmosphere. At even higher temperatures, both
isotopes are enriched and the isotopic anomaly vanishes as the helium
abundance rises (see also Vauclair et al. \cite{vauclair74}).

A similar scenario {  might explain} the more compact sdB
stars. Focusing on the upper helium sequence in Fig.~\ref{fig:nhevsteff} one
can see that the helium abundance is scattering around $\log{y}\simeq-2.5$
for temperatures below $31\,000\,{\rm K}$. $^{3}$He is enriched at the
border region. For temperatures higher than $\simeq31\,000\,{\rm K}$ the
helium abundance is rising.

Finally, radiatively driven stellar wind mass loss might play a role ({ 
see, for example}, Babel \cite{babel96} and references therein), but Hu et
al. (\cite{hu11}) derived upper limits for this mass loss to be of the order
of $10^{-15}\,M_{\rm \odot}\,{\rm yr^{-1}}$ to allow sdBs to pulsate.
However, even weak and fractionated winds might already affect the
abundances in the atmospheres of sdB stars (Unglaub \cite{unglaub08}).

\section{Summary}\label{sec:summary}

Atmospheric parameters and helium abundances of 44 bright sdBs have been
determined. We {  have shown} that effective temperatures and surface
gravities can be derived from high resolution echelle spectra with
sufficient accuracy. Systematic uncertainties have been determined by
comparing the parameters derived from the high resolution data with the ones
derived from medium resolution spectra. Most stars are core helium-burning,
but some sdBs are already in the shell helium-burning phase.

Helium abundances have been measured with high accuracy. Besides the known
correlation of helium abundance with temperature, two distinct sequences in
helium abundance have been confirmed. The reasons for both the { 
increasing} helium abundance with temperature and the bimodal distribution
{  have been discussed, but} we are left without a strong conclusion.

Significant isotopic shifts of helium lines due to an enrichment in $^{3}$He
have been found in the spectra of 8 sdBs. Most of these stars cluster in a
small temperature range between $27\,000\,{\rm K}$ and $31\,000\,{\rm K}$
very similar to the known $^{3}$He-rich main sequence B stars, which cluster
at somewhat lower temperatures. This phenomenon is most probably related to
diffusion processes in the atmosphere.

\begin{acknowledgements}

S.~G. was supported by the Deutsche Forschungsgemeinschaft under grant 
He~1356/49-1. D.~K. thanks the University of the Western Cape and the South African National Research Foundation for financial support. We thank R. A. Saffer for sharing his data with us as well as N. Hambly and H. McGillivray for their contributions to the EC survey. We also thank S.~J. O'Toole and S. Moehler for fruitful discussions. {  Finally, we want to thank the referee S. Vennes for helpful comments and suggestions.}
Based on observations at the La Silla Observatory of the 
European Southern Observatory for programmes number 073.D-0495(A),
  074.B-0455(A), 076.D-0355(A), 077.D-0515(A) and 078.D-0098(A). Based on
  observations with the William Herschel Telescope and the Isaac Newton
  Telescope operated both by the Isaac Newton Group at the Observatorio del
  Roque de los Muchachos of the Instituto de Astrofisica de Canarias on the
  island of La Palma, Spain. This paper uses observations made at the South
  African Astronomical Observatory (SAAO). Based on observations collected
  at the Centro Astron\'omico Hispano Alem\'an (CAHA) at Calar Alto,
  operated jointly by the Max-Planck Institut f\"ur Astronomie and the
  Instituto de
   Astrof\'isica de Andaluc\'ia (CSIC).

\end{acknowledgements}

\begin{appendix}

\section{Comparison of atmospheric parameters with the literature}\label{sec:lit}

Atmospheric parameters of 20 {  of the} sdBs analysed here {  have also
been} found in the literature. All parameters derived from model fits to the
spectral lines are listed in Table~\ref{tab:lit}. Given the fact that
different methods, models and data were used to determine these parameters,
our new results are in general agreement with the literature values. The
lower temperature ($-2600\,{\rm K}$) Saffer et al. (\cite{saffer94}) derived
for $[$CW83$]$\,1758$+$36 may be more accurate, because our analysis is
based on only one FOCES spectrum of mediocre quality, while the spectra used
by Saffer et al. (\cite{saffer94}) cover the Balmer jump. We reanalysed the
spectrum used by Saffer et al. (\cite{saffer94}) with our method and
confirmed the parameters given in this study (Saffer refit).

The atmospheric parameters of the bright sdB, PG\,0342+026, have been
determined from its spectral energy distribution by Lamontagne et al.
(\cite{lamontagne87}), Theissen et al. (\cite{theissen95}) and Aznar
Cuadrado \& Jeffery (\cite{aznar01}). Given the higher uncertainties of this
method the derived parameters are consistent with our results within the
error limits.

\begin{table*}
\caption{Atmospheric parameters from the literature. The first lines refer to the results derived in this paper.}
\begin{center}
\label{tab:lit}
\begin{tabular}{llllllll}
\hline
\noalign{\smallskip}
System & other names & $T_{\rm eff}$ & $\log{g}$ & $\log{y}$ & reference \\
       & & [K] &  &  & \\ 
\noalign{\smallskip}
\hline
\noalign{\smallskip}
$[$CW83$]$\,0512$-$08   &         & 38400 & 5.77 & -0.73 & \\
                        &         & 38000 & 5.6  &       & Edelmann et al. \cite{edelmann01} \\
$[$CW83$]$\,1758$+$36   & PG\,1758+364 & 34600 & 5.79 & -1.51 & \\
                        &         & 32100 & 5.91 & -1.82 & Saffer et al. \cite{saffer94} \\
                        &         & 32500 & 5.73 & -1.85 & Saffer refit \\
EC\,03591$-$3232        &  CD\,$-$32\,1567       & 28000 & 5.55 & -2.03 & \\
                        &         & $27000\pm1300$ & $5.36\pm0.19$ & $-1.63\pm0.21$  & Vennes et al.  \cite{vennes11} \\
                        &         & $30490\pm240$  & $5.71\pm0.05$ & $-1.92\pm0.05$  & N\'emeth et al. \cite{nemeth12} \\
EC\,14248$-$2647        &         & 31400 & 5.56 & -1.64 & \\
                        &         & $31880\pm300$ & $5.70\pm0.07$ & $-1.71\pm0.08$   & N\'emeth et al. \cite{nemeth12} \\
EC\,22081$-$1916        &         & 31100 & 4.77 & -1.97 & see also Geier et al. \cite{geier11a} \\
Feige\,38               & PG\,1114+072     & 30600 & 5.83 & -2.37 &  \\
                        &         & 29800 & 5.81 & -2.22 & Saffer et al. \cite{saffer94} \\
Feige\,65               & PG\,1233+426  & 26200 & 5.31 & -2.75 & \\
                        &         & 26500 & 5.60 & -2.3  & Saffer et al. \cite{saffer94} \\
LB\,1516                &         & 25200 & 5.41 & -2.78 & \\
                        &         & 26300 & 5.7  & -2.5  & Heber \cite{heber86} \\
                        &         & 26100 & 5.4  &       & Chayer et al. \cite{chayer06} \\
PB\,5333                &         & 40600 & 5.96 & -2.62 & \\
                        &         & 37900 & 5.81 & -2.70 & Saffer et al. \cite{saffer94} \\
PG\,0342$+$026          &         & 26000 & 5.59 & -2.69 &   \\
                        &         & 26200 & 5.67 & -2.4  & Saffer et al. \cite{saffer94} \\
PG\,0909$+$164          &         & 35300 & 5.33 & -2.76 & \\
                        &         & 35400 & 5.64 & -2.70 & Saffer priv. comm., Maxted et al. \cite{maxted01} \\
PG\,0909$+$276          &         & 35500 & 6.09 & -1.00 & \\
                        &         & 35400 & 6.02 & -0.92 & Saffer et al. \cite{saffer94} \\
PG\,1303$+$097          &         & 29800 & 5.83 & -2.17 & \\
                        &         & 30300 & 5.76 & -1.96 & Saffer priv. comm., Maxted et al. \cite{maxted01} \\
PG\,1505$+$074          &         & 37100 & 5.39 & -2.69 & \\
                        &         & 37100 & 5.42 & -3.1 & NLTE, Maxted et al. \cite{maxted01} \\
PG\,1616$+$144          &         & 37300 & 5.95 & -1.26 & \\
                        &         & 36500 & 6.02 & -1.51 & NLTE, Maxted et al. \cite{maxted01} \\
PG\,1710$+$490          &         & 30600 & 5.66 & -2.43  &  \\
                        &         & 29900 & 5.74 & -2.22 & Saffer et al. \cite{saffer94} \\
                        &         & 30300 & 5.7  &       & Chayer et al. \cite{chayer06} \\
PG\,2314$+$076          &         & 27200 & 5.65 & $<-4.0$ & \\
                        &         & 28600 & 5.75 & $<-4.0$ & Saffer priv. comm. refit \\
PG\,2349$+$002          &         & 28000 & 5.73 & -3.45 & \\
                        &         & 29300 & 5.77 &       & Saffer et al. \cite{saffer94} \\
PHL\,457                & GD\,1110  & 26500 & 5.38 & -2.54 & \\
                        &         & 25000 & 5.3 & -2.44  & Heber \cite{heber86} \\
                        &         & 28200 & 5.5 & -2.5   & NLTE, Blanchette et al. \cite{blanchette08} \\
                        &         & 29300 & 5.6 & -2.4   & LTE, Blanchette et al. \cite{blanchette08} \\
SB\,815                 & CD\,$-$35\,15910  & 27000 & 5.32 & -2.90 & \\
                        &         & 28800 & 5.4 & -2.44  & Heber et al. \cite{heber84} \\
                        &         & $28390\pm300$ & $5.39\pm0.04$ & $-3.07\pm0.2$  & N\'emeth et al. \cite{nemeth12} \\
\noalign{\smallskip}
\hline
\end{tabular}
\end{center}
\end{table*}

\end{appendix}

\end{document}